# SMASH: Physics-guided Reconstruction of Collisions from Videos

Aron Monszpart[1]   Nils Thuerey[2]   Niloy J. Mitra[1]
[1]University College London   [2]Technical University of Munich

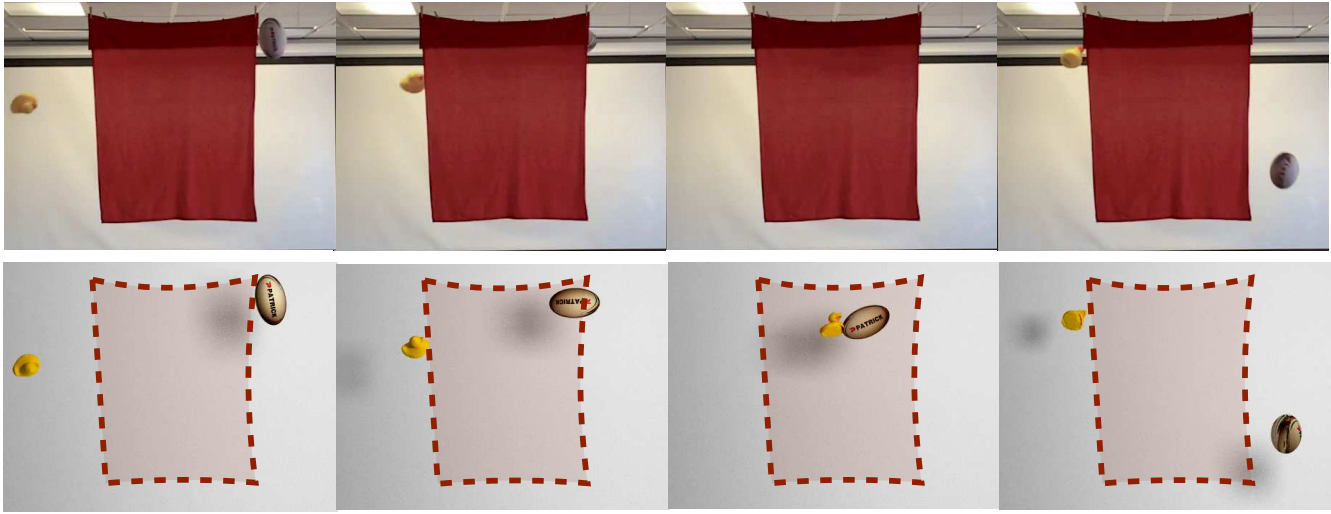

**Figure 1:** *Starting from an input video of a collision sequence behind a curtain (top),* SMASH *reconstructs an accurate physically valid collision (bottom) using laws of rigid body physics for regularization. Note the reconstructed spin (i.e., angular velocity) of the objects.*

## Abstract

Collision sequences are commonly used in games and entertainment to add drama and excitement. Authoring even two body collisions in the real world can be difficult, as one has to get timing and the object trajectories to be correctly synchronized. After tedious trial-and-error iterations, when objects can actually be made to collide, then they are difficult to capture in 3D. In contrast, synthetically generating plausible collisions is difficult as it requires adjusting different collision parameters (*e.g.,* object mass ratio, coefficient of restitution, *etc.*) and appropriate initial parameters. We present SMASH to directly *read off* appropriate collision parameters directly from raw input video recordings. Technically we enable this by utilizing laws of rigid body collision to regularize the problem of lifting 2D trajectories to a physically valid 3D reconstruction of the collision. The reconstructed sequences can then be modified and combined to easily author novel and plausible collisions. We evaluate our system on a range of synthetic scenes and demonstrate the effectiveness of our method by accurately reconstructing several complex real world collision events.

**Concepts:** •**Computing methodologies** → **Computer graphics; Physical simulation;** *Simulation evaluation; Motion processing;*

## 1  Introduction

Collisions capture suspense, build anticipation, and pack drama. Naturally, they remain an integral part of movies, games, and entertainment. Creating a good real-world collision sequence involving multiple objects, however, is difficult. While the act of smashing two objects into each other so that they collide in a certain way is already non-trivial, the setup quickly becomes unmanageable when additional colliding objects are to be collided, or adjustments are required to the object trajectories. Such changes can easily require many further iterations and recordings, and become a tedious trial-and-error process. Moreover, trying out multiple collision iterations with expensive or fragile objects may not even be a realistic option.

Accurately capturing real-world collision sequences poses further challenges. On one hand, such sequences necessitate high to very high framerate capture, thus making state-of-the-art methods like Kinect Fusion, *etc.,* unsuitable candidates for 3D acquisition. On the other hand, even high-framerate video data only provides partial 2D information, both in space and in time (see Figure 2). A fundamental problem arises due to unavoidable occlusions near collision instances, which prevents direct observation of the actual collision processes in any acquisition setup.

While the physics of object collisions is a challenging problem in itself, well-developed high-level models exist to reduce its complexity. One widely used assumption is that of infinite object stiffness, *i.e.,* ideally rigid motions. Whereas such rigid body simulations are widely used in games and movies, the task of setting up a collision with the right initial conditions remains tough: many parameters, such as velocities, mass ratios, and coefficient of restitution, have to be correctly guessed and adjusted. Given the nonlinear nature of the underlying physics, such a rigging up of a desirable sequence is problematic and typically requires extensive prior experience. Further, physical parameters (*e.g.,* coefficient of restitution) may not be readily available for the object pair at hand. Finally, such an approach only enables forward simulations.

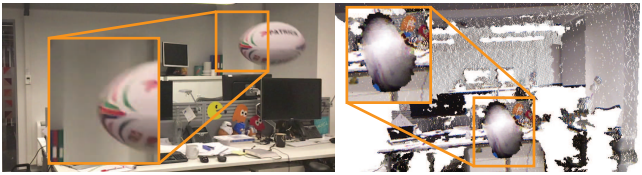

**Figure 2:** *Accurate acquisition of collision sequences is difficult as objects move fast and are typically occluded around collision time: a video captured by a smartphone (left) showing motion blur, and an RGBD scan using Kinect (right) that is noisy and partial.*

In this work, we propose to marry the benefits of the above setups. The user records collisions between pairs of objects simply using a high framerate video (a smartphone in our setup). The video data, however, lacks depth information and is noisy. We first automatically extract a set of candidate 3D positions from the video, and semi-automatically initialize a sparse set of orientations. We then formulate an optimization using regularization derived from conservation laws of rigid body collisions to reconstruct space-time trajectories of the participating objects. This step utilizes the original 3D models of the colliding objects, assumed to be captured beforehand. As output, we can directly *read off* physical collision parameters from the reconstruction that can readily be incorporated into existing physics engines to recreate and reauthor modifications of the recorded collisions. For example, in Figure 1, we show the replay of the reconstructed collision sequence happening behind the curtain. Note that we obtain high-quality reconstruction results even when the most relevant section (in time) of the original collision remains fully occluded in the input video.

We evaluate our method both on a range of synthetic evaluation setups and complex real-world collision examples, and generate new collision sequences using an easy authoring workflow. In summary, we propose an algorithm to reconstruct physically valid collisions from an input video, and perform collision analysis without access to exact object geometry. The estimated collision parameters can then be used to author new collisions sequences.

## 2 Related work

**Rigid body simulation** has a long and successful history within Computer Graphics. Starting with the early works in animation [Armstrong and Green 1985; Baraff 1990], they are now widely used in all forms of computer animation. A good overview can be found in the book by Eberly [2010]. While the forward simulation problem is far from trivial, it is well studied. We are targeting an inverse problem, and in the following we restrict discussion to the relevant works.

The first methods to edit or modify simulations proposed spacetime methods to compute optimal motions in constrained systems [Witkin and Kass 1988; Liu et al. 1994]. Variants with neural networks [Grzeszczuk et al. 1998] and genetic algorithms were proposed [Tang et al. 1995] to reduce the complexity or synthesize directable motions. A randomized approach to re-construct motions for target configurations of rigid bodies was proposed by Chenney et al. [2000], while an interactive variant was developed by Popović et al. [2000]. While these algorithms offer varying levels of control and parameter estimation for rigid bodies, they focus on virtual situations in a simulator. In contrast, we propose a method that works robustly with only sparse, unreliable data from a real world source.

Areas that would be highly interesting, but which we have currently not taken into account are deformable objects [Terzopoulos et al. 1987; Martin et al. 2011] and fracture [Müller et al. 2001; Su et al. 2009]. In the following, we assume that the observed objects do not deform or change topology.

Others have used the complexity of collision events to randomize the solution space, in this way giving control over the outcome of a simulation [Twigg and James 2007]. Recently, Smith et al. [2012] investigated the even tougher case of multiple, simultaneous collisions. While we focus on the two body case, these papers highlight the complexities of collisions, which are amplified for imperfect real-world objects. A good general survey of collision modeling is the report by Gilardi and Sharf [2002].

An interesting general direction of research investigates the retrieval of modeling equations from data by employing learning techniques [Bongard and Lipson 2007]. While our method is not directly using machine learning, it represents a *data-driven* approach to recover complex real-world phenomena. In this context, we demonstrate that suitable physical models are powerful regularizers for underdetermined problems.

Conotter et al. [2012] analyze the kinematics of the ballistic motion of a single object in videos in the digital forensics context. Salzmann and Urtasun [2011] explain the change in the linear motion of objects by solving for a sparse set of forces, applying their dynamics motivated regularization in video based tracking. We focus on acquiring the intrinsic physical properties of the objects to guide our motion estimates taking both linear and angular motion into account.

**Geometry acquisition** has been simplified with affordable and portable scanning options. Hence, in recent years, significant efforts have gone towards capturing and parsing 3D scenes. Approaches include using classifiers on scene objects [Schlecht and Barnard 2009; Xiong and Huber 2010; Anand et al. 2013; Koppula et al. 2011; Silberman et al. 2012], interactive 3D modeling from raw RGBD scans [Shao et al. 2012], interleaving segmentation and classification [Nan et al. 2012], unsupervised algorithms to identify and consolidate scans [Kim et al. 2012; Mattausch et al. 2014], dynamic reconstruction [Mitra et al. 2007], proxy geometry based scene understanding [Lafarge and Alliez 2013; Monszpart et al. 2015], or studying the spatial layout of scenes [Gupta et al. 2010; Lee et al. 2010; Hartley et al. 2012].

Physical constraints have also been used for scene understanding: for example, [Jia et al. 2013; Jiang and Xiao 2013; Zheng et al. 2013; Shao* and Monszpart* et al. 2014] consider local and global physical stability to predict occluded geometry in scenes. These methods, however, primarily focus on static scenes. In case of dynamic scenes, even state-of-the-art RGBD based systems [Newcombe et al. 2015; Dou et al. 2015] struggle with highly dynamic motion and occlusion. Capturing 3D geometry of dynamic scenes remains very challenging. For dynamic scenarios, inverse methods with physics priors have also been used to capture other phenomena such as liquids [Wang et al. 2009] or smokes [Gregson et al. 2014].

While impressive results have been demonstrated in case of template-based solutions for human faces, hair, body, *etc.*, they focus on application specific contexts where object behavior and dynamics can be captured and learned in a training phase. In contrast, we focus on reconstructing 3D geometry of collisions directly from raw videos. Note that due to the nature of the problem, collisions happen prohibitively fast for accurate capture. Further, object parts are severely occluded near the actual collision. We demonstrate, that motion states of the objects as well as internal physical properties can be estimated by analysing the objects' pre- and post-collision trajectories away from the time of collision.

# 3 Formulation

Our goal is to acquire pairwise relevant parameters of object collisions simply based on an input video and reconstruct a physically valid motion. As output we produce a space-time recording of the collision event that can be re-used in a variety of ways: to set up new collisions that behave faithfully to the original recording, to introduce new objects, or even to author complex interactions between objects by combining multiple recordings.

We first briefly review rigid body dynamics (Section 3.1) as the governing equations provide important building blocks for our optimization. We then formulate the problem (Section 3.2) and identify the appropriate conservation laws to constrain solutions to the space of physically-plausible ones (Section 3.3 - 3.6). We solve the problem in an energy-minimization framework using the above components. In the following, we use bold lower-case letters for 3D and 4D vectors (*e.g.*, $\mathbf{p}$), and reserve bold upper case letters for matrices (*e.g.*, $\mathbf{R}$), see Table 1. All our measurements are in world space units.

## 3.1 Rigid Bodies

By using a rigid body model we represent an object by its motion around its center of mass $\mathbf{p}$. In addition to $\mathbf{p}$, each body has an orientation, represented as a unit quaternion $\mathbf{q}$ (we denote the equivalent rotation matrix by $\mathbf{R_q}$). Due to rigidity, we only need to consider linear and angular velocity ($\mathbf{v}$ and $\boldsymbol{\omega}$, respectively). A body further has a mass and a moment of inertia (calculated for a reference orientation), denoted by $m$ and $\mathbf{I}_0$, respectively. While $\mathbf{p}, \mathbf{q}, \mathbf{v}$, and $\boldsymbol{\omega}$ change over time, $m$ and $\mathbf{I}_0$ are assumed to be constant. For notational simplicity, we keep the time dependence, *e.g.*, $\mathbf{p}(t)$, as much as possible implicit and simply write $\mathbf{p}$. The instantaneous velocity of any point $\mathbf{x}$ on a rotating body is given by $\mathbf{v} + \boldsymbol{\omega} \times (\mathbf{x} - \mathbf{p})$.

Frictionless collisions between two objects are typically modelled with a scalar impulse $j$ acting along a collision normal $\mathbf{n}$. For two bodies $a$ and $b$, we denote their respective variables with a superscript, *e.g.*, $\mathbf{p}^{\text{a}}$ is the center of mass of object $a$. We will use $^\text{c}$ for variables related to the collision event. The impulse changes pre-collision velocities into a set of post-collision velocities (denoted with superscript $^{\text{pre}}$ and $^{\text{post}}$, respectively) such that the scalar relative velocity $v_r = \mathbf{v}_r \cdot \mathbf{n}$ at the point of collision satisfies $v_r^{\text{post}}/v_r^{\text{pre}} = -c$. Here, $c$ is the coefficient of restitution, which is related to the amount of energy that is transferred into a reversal of the object's motion. The remainder is *lost* for the simulation and dissipated into heat, sound, or work to deform the internal structure of the objects.

In rigid body solvers we typically use a chosen value for $c$ to compute $j$, and with it the post-collision velocities. The impulse acts anti-symmetrically to conserve momentum, both linear:

$$\mathbf{v}^{\text{post,a}} = \mathbf{v}^{\text{pre,a}} + j\mathbf{n}/m^{\text{a}},$$
$$\mathbf{v}^{\text{post,b}} = \mathbf{v}^{\text{pre,b}} - j\mathbf{n}/m^{\text{b}}, \qquad (1)$$

and angular, where the instantaneous change of the angular momentum $\mathbf{k} = \mathbf{I}\boldsymbol{\omega}$ is

$$\mathbf{k}^{\text{post,a}} = \mathbf{k}^{\text{pre,a}} + ((\mathbf{x}^{\text{c}} - \mathbf{p}^{\text{a}}) \times j\mathbf{n}),$$
$$\mathbf{k}^{\text{post,b}} = \mathbf{k}^{\text{pre,b}} - ((\mathbf{x}^{\text{c}} - \mathbf{p}^{\text{b}}) \times j\mathbf{n}). \qquad (2)$$

## 3.2 Problem Statement

In contrast to the typical forward simulation above, our goal is to retrieve the physical parameters directly from a real collision of two objects. This would normally involve a large amount of tedious manual work or complicated capturing setups. We now explain our *inverse approach* to retrieve the physical parameters based on observations of the objects' trajectories. It should be pointed out here that computing absolute quantities (*e.g.*, mass) is not possible without access to a reference measurement. We will not be able to compute an absolute position of the objects on earth from a single video input, and, correspondingly, we cannot compute their absolute masses. However, we can retrieve relative quantities, *i.e.*, the relative positions, and the relative mass of the objects. We neglect all other external effects (*e.g.*, aerodynamic drag, friction) at work except for a known gravitational acceleration with magnitude $g = 9.81 m/s^2$.

|  | inputs | for each … |
|---|---|---|
| $g$ | gravitational acceleration | global |
| $f_{\text{fps}}$ | video frame rate | recording |
| $\mathbf{P}$ | camera projection matrix | recording |
| $\mathbf{q}^{\text{obs,a}}, \mathbf{q}^{\text{obs,b}}$ | orientation inputs | annotation |
|  | *derived quantities* | *for each …* |
| $\mathbf{p}^{\text{a}}, \mathbf{p}^{\text{b}}$ | object position | cont. time $t$ |
| $\mathbf{q}^{\text{a}}, \mathbf{q}^{\text{b}}$ | object orientation | cont. time $t$ |
| $\mathbf{v}^{\text{a}}, \mathbf{v}^{\text{b}}$ | linear velocity | cont. time $t$ |
| $\boldsymbol{\omega}^{\text{a}}, \boldsymbol{\omega}^{\text{b}}$ | angular velocity (world) | cont. time $t$ |
| $\mathbf{I}_0^{\text{a}}, \mathbf{I}_0^{\text{b}}$ | inertia tensor (local) | object |
| $\mathbf{p}^{2\text{d,a}}, \mathbf{p}^{2\text{d,a}}$ | image-space position | frame (discr. time) |
| $s^{2\text{d}}$ | bounding circle diameter | frame (discr. time) |
| $d^{\text{a}}, d^{\text{b}}$ | approximate depth | frame (discr. time) |
|  | **unknowns** | *(number)* |
| $\beta_x, \beta_{y1}$ | gravity rotation angles | recording (2) |
| $\mathbf{b}_3^{\text{a}}, \mathbf{b}_3^{\text{b}}$ | parabola translation (collision pos.) | object (2×3) |
| $\mathbf{q}^{\text{c,a}}, \mathbf{q}^{\text{c,b}}$ | orientation at collision time | object (2×4) |
| $\mathbf{k}^{\text{pre,a}}, \mathbf{k}^{\text{pre,b}}, \mathbf{k}^{\text{post,a}}, \mathbf{k}^{\text{post,b}}$ | angular momentum | segment (4×3) |
| $b_1^{\text{pre,a}}, b_1^{\text{pre,b}}, b_1^{\text{post,a}}, b_1^{\text{post,b}}$ | parabola linear coefficients (x coord.) | segment (4×1) |
| $b_2^{\text{pre,a}}, b_2^{\text{pre,b}}, b_2^{\text{post,a}}, b_2^{\text{post,b}}$ | parabola linear coefficients (y coord.) | segment (4×1) |
| $\beta_{y0}^{\text{pre,a}}, \beta_{y0}^{\text{pre,b}}, \beta_{y0}^{\text{post,a}}, \beta_{y0}^{\text{post,b}}$ | parabola rotation angles | segment (4×1) |
| $\mathbf{x}^{\text{c}}$ | collision point (world) | collision (1x3) |
| $j\mathbf{n}$ | impulse times collision normal | collision (1x3) |
| $t^{\text{c}}$ | collision time | collision (1) |
| $c$ | coefficient of restitution | collision (1) |
| $m^{\text{b,a}}$ | mass ratio | collision (1) |

**Table 1:** *Notation table.*

Interestingly, the actual shape of the objects does not play a direct role. Only a related quantity, the distribution of mass is important in the form of the moment of inertia. We diagonalize the moment of inertia for all objects by pre-aligning the initial configuration along their principal axes. Thus, we represent an object's moment of inertia with three degrees of freedom along the diagonal of $\mathbf{I}_0$,

$$\mathbf{I}_0 = \frac{1}{m} diag \begin{pmatrix} i_{0,0} & i_{1,1} & i_{2,2} \end{pmatrix}, \qquad (3)$$

$$\mathbf{I} = \mathbf{R_q} \mathbf{I}_0 \mathbf{R_q}^{-1}. \qquad (4)$$

Starting from an input video, we first automatically extract a dense set of image-space centroids with associated depth values and semi-automatically generate a sparse set of orientations. While we formulate the problem in 3D, we will later on minimize the projected error in screen space.

Our goal is to compute: a parametrization of the trajectories of the two bodies $a$ and $b$, their relative mass, their pre- and post-collision velocities (linear as well as angular), the time of collision, the collision impulse, and the coefficient of restitution. These quantities should match the observed data from the input video as well as possible. We will describe our approach to incorporate these unknowns in our solution in this order. As the image data is potentially unreliable and noisy, we do not rely purely on the corresponding data-terms in our calculations, but use the physical laws as regularizers. One of our central contributions is identifying and incorporating the relevant physical constraints.

### 3.3 Center of Mass Trajectories

A well known fact is that the trajectory of the center of mass of an object experiencing a constant acceleration is given by a parabola. This description becomes invalid at the time of the collision $t^c$, but it is an excellent model for the trajectories before and after, as long as effects such as aerodynamic drag are negligible. For the two objects, with pre- and post-collision trajectories, we thus parametrize and extract four 3D parabolas from the inputs. We parametrize each parabola in 3D space with

$$\mathbf{p}(t) = \mathbf{R} \begin{pmatrix} b_1(t - t^c) \\ -\frac{g}{2}(t - t^c)^2 + b_2(t - t^c) \\ 0 \end{pmatrix} + \mathbf{b}_3, \quad (5)$$

where $b_1, b_2$ parametrize the curve over time, the vector $\mathbf{b}_3$ determines its offset, and the rotation matrix $\mathbf{R}$ its orientation. We shift the curve in time by the collision time $t^c$, so that $\mathbf{p}(t^c) = \mathbf{b}_3$.

Moreover, the different parabolas can only rotate about the axis of gravity, which, without loss of generality, we assume to be the y-axis in our formulation. We encode this premise as a global rotation with two degrees of freedom shared by all four parabolas and individual rotations around the gravity direction. As a consequence, we construct a rotation matrix $\mathbf{R}$ as a sequence of Euler angle rotations. We chose the proper Euler angle representation Y,X,Y, the angles of which we denote with $\beta_{y0}, \beta_x$, and $\beta_{y1}$, respectively. We denote the resulting rotation matrix by $\mathbf{R}_{\beta_{y0}, \beta_x, \beta_{y1}}$. Note that $\beta_{y0}$ is different for every parabola, while the other two are shared, rotating the common gravity vector.

Further, the two parabolas for an object coincide in one point at the time of collision, allowing us to express both parabolas uniquely with a single $\mathbf{b}_3$ offset. In total, that leaves us with two global angles, four individual angles, two offsets, and four times $b_1, b_2$ as unknowns. The geometric setup of the four parabolas is illustrated in Figure 3.

We directly estimate the linear velocity from these curves with the temporal derivative of Eq. (5). The most important velocity in our setting is the velocity at collision time $t^c$:

$$\mathbf{v}(t^c) = \mathbf{R} \begin{pmatrix} b_1 & b_2 - gt^c & 0 \end{pmatrix}^T \quad (6)$$

We now describe the angular aspects of our motion modeling.

### 3.4 Orientation and Angular Momentum

While the angular velocity of a rigid body can change without external forces, its angular momentum remains constant (see [Kleppner and Kolenkow 2013]). Thus, for each of the four trajectories, we calculate an angular momentum $\mathbf{k}$ that best explains the orientations observed over time. Angular momentum and velocity relate to each other with:

$$\boldsymbol{\omega}(t) = \mathbf{I}^{-1}\mathbf{k}(t) = \mathbf{R}_\mathbf{q}(t)\mathbf{I}_0^{-1}\mathbf{R}_\mathbf{q}^{-1}(t)\mathbf{k}(t) \quad (7)$$

Once we know the angular momentum at time $t$, we can compute the corresponding angular velocity $\boldsymbol{\omega}$, and use it to integrate an orientation forward in time. With $\boldsymbol{\omega}$ as the imaginary part of a quaternion, an Euler step is given by:

$$\mathbf{q}(t + \Delta t) = \mathbf{q}(t) + \frac{\Delta t}{2} \begin{pmatrix} 0 \\ \boldsymbol{\omega} \end{pmatrix} \otimes \mathbf{q}(t) \quad (8)$$

where $\otimes$ denotes the quaternion product, and we normalize $\mathbf{q}(t + \Delta t)$ after each step.

Similar to the shared offset of the parabolas, we introduce an orientation $\mathbf{q}^c$ for each body at collision time that is shared by the two parabola segments. Our goal is not to use one of the input orientations directly, due to their potential unreliability. Instead we find an improved orientation that, together with an angular momentum $\mathbf{k}$, explains the observed orientations as well as possible. Starting with $\mathbf{q}^c$ at collision time $t^c$ we integrate this orientation backward or forward to time $t$ with explicit integration. We perform a series of integration steps with Eq. (8) from $t^c$ to a given time $t$. This yields the orientation of body $a$ or $b$ at any instance in time based on the estimated orientation at collision time. We will later on use this step to calculate the estimated orientation at different times, based on our current estimate for the collision orientation and object motion. Thus it will later on serve to guide the optimization towards the input orientations.

### 3.5 Energy Limiting

When gravity is the only external force present, the sum of kinetic and potential energy (for height $h$) remains constant as:

$$E = \tfrac{1}{2}m\,\mathbf{v}^T\mathbf{v} + \tfrac{1}{2}\boldsymbol{\omega}^T\mathbf{I}\boldsymbol{\omega} + mgh. \quad (9)$$

Upon collision, a portion of this energy dissipates. This is typically modeled with the aforementioned coefficient of restitution. As it is one of the central parameters that we want to retrieve, we cannot directly impose constraints conserving the energy in our system. However, assuming the collision is an instantaneous event, we can ensure that the post-collision energy is less or equal to the pre-collision energy. As position and orientation do not change at the time of collision, the potential energy remains the same. Thus, our formulation focuses on the kinetic energy, and we will later on ensure that it does not increase after the collision.

### 3.6 Conservation of Momentum

Throughout the objects' trajectories, angular momentum is conserved in the absence of external torques. Linear momentum is only conserved in the direction orthogonal to gravity. Instantaneous rigid body collisions fully preserve both total linear and total angular momentum of the objects involved. As such, the equations for an impulse-based collision are ideal to couple the four separate trajectories at the time of collision. We have already exploited the fact that the two parabolas for each object have to connect in the respective collision positions, additionally, the conservation laws for momentum enable the coupling of the two objects' trajectories.

Conservation of linear momentum means that the sum of linear momenta before and after collision has to be equal. Since we cannot recover the absolute mass, we set the mass of body $a$ to 1.0, and introduce a mass ratio $m^{b,a} = m^b/m^a$, (we will revisit the case of a potentially infinitely heavy object in Section 5.2). Formulating the equations in terms of $m^{b,a}$ yields

$$\mathbf{v}^{\text{pre},a}(t^c) + \mathbf{v}^{\text{pre},b}(t^c)m^{b,a} = \mathbf{v}^{\text{post},a}(t^c) + \mathbf{v}^{\text{post},b}(t^c)m^{b,a}. \quad (10)$$

As the velocities are determined from the parabolas, this equation ensures that these curves do not create or lose linear momentum, and effectively couple the planes of the four parabolas.

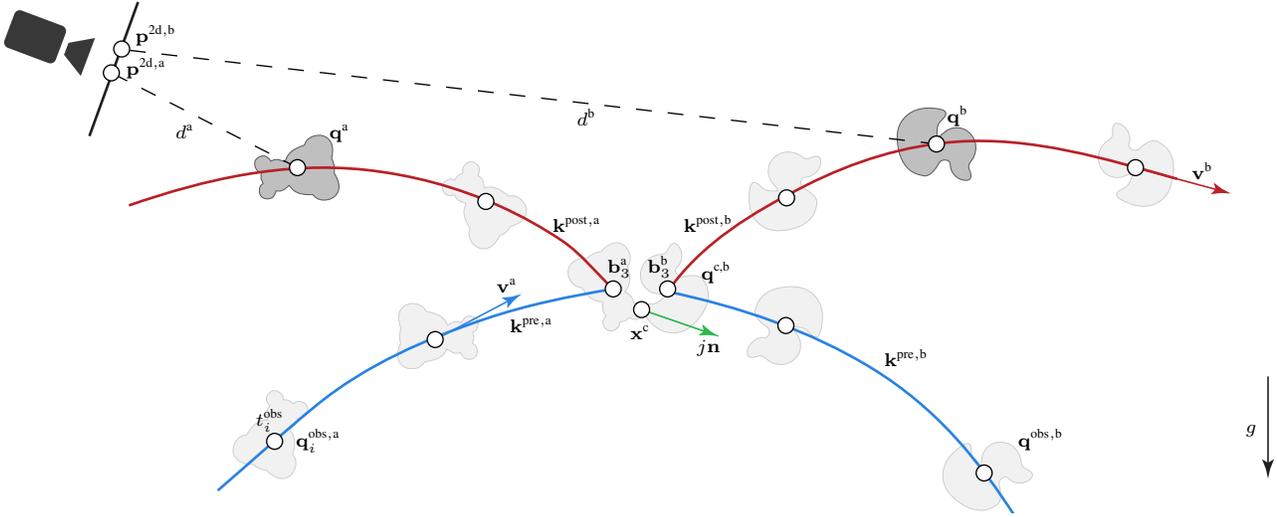

**Figure 3:** *Modeling two-body collisions: We extract dense position estimates from a video input and augment it with sparse semi-automatically generated orientations. These observations are linked with laws of physics: the individual trajectories should be parabolas as we assume the bodies to have ballistic motion, and they are coupled at the (unknown) collision point based on conservation laws.*

We similarly ensure that the angular momentum we compute does not violate conservation of angular momentum. For the angular momenta, we have to consider their sum, and additionally the instantaneous angular momentum around the origin:

$$\mathbf{p}^a \times \mathbf{v}^{\text{pre,a}} + \mathbf{k}^{\text{pre,a}} + m^{b,a}\mathbf{p}^b \times \mathbf{v}^{\text{pre,b}} + \mathbf{k}^{\text{pre,b}} =$$
$$\mathbf{p}^a \times \mathbf{v}^{\text{post,a}} + \mathbf{k}^{\text{post,a}} + m^{b,a}\mathbf{p}^b \times \mathbf{v}^{\text{post,b}} + \mathbf{k}^{\text{post,b}} \quad (11)$$

Note, that we use $m^{b,a}$ to compute $\mathbf{I}_0^b$ and $\mathbf{k}^b$ using Eq. (3). All quantities in Eq. (11) are evaluated at collision time $t^c$. The two equations (10) and (11) constrain our solution space to a physically plausible one, and provide a first coupling between objects $a$ and $b$. We improve this coupling with the help of the impulse equations from Section 3.1. The equations for linear impulse (Eq. (1)) express the exchange of momentum between the two bodies relative to their mass:

$$\mathbf{v}^{\text{post,a}}(t^c) = \mathbf{v}^{\text{pre,a}}(t^c) + \frac{j\mathbf{n}}{1},$$
$$\mathbf{v}^{\text{post,b}}(t^c) = \mathbf{v}^{\text{pre,b}}(t^c) - \frac{j\mathbf{n}}{m^{b,a}}. \quad (12)$$

This equation turns out to be crucial for retrieving the mass ratio $m^{b,a}$ during the combined optimization, which we will detail below.

The angular impulse equations allow us to enforce a shared collision point $\mathbf{x}^c$. Eq. (2) connects the angular momenta of the two bodies via the relative offset of the collision point:

$$\mathbf{k}^{\text{post,a}} = \mathbf{k}^{\text{pre,a}} + ((\mathbf{x}^c - \mathbf{b}_3^a) \times j\mathbf{n})$$
$$\mathbf{k}^{\text{post,b}} = \mathbf{k}^{\text{pre,b}} - ((\mathbf{x}^c - \mathbf{b}_3^b) \times j\mathbf{n}). \quad (13)$$

We complete the set of necessary equations by computing the relative velocities at the collision point to compute the coefficient of restitution. We only need to consider the velocity components along the collision normal direction:

$$c = -\frac{\left(\hat{\mathbf{v}}^{\text{post,a}} - \hat{\mathbf{v}}^{\text{post,b}}\right) \cdot \mathbf{n}}{\left(\hat{\mathbf{v}}^{\text{pre,a}} - \hat{\mathbf{v}}^{\text{pre,b}}\right) \cdot \mathbf{n}}, \quad (14)$$

with $\hat{\mathbf{v}} = \mathbf{v}(t^c) + \boldsymbol{\omega}(t^c) \times (\mathbf{x}^c - \mathbf{p}(t^c))$ and $\boldsymbol{\omega}(t^c) = \mathbf{I}^{-1}(t^c)\mathbf{k}(t^c)$. With these equations at hand we now formulate an optimization problem to compute the unknown quantities as described next.

## 4 Method

Our method proceeds in three stages: First, we estimate center of mass positions based on the input video. Second, we initialize orientations using a differential raytracer, with user input for ambiguous situations. Third, in the main step, we take these position and orientation estimates to retrieve the physical parameters of the rigid body motion and collision. For the initialization, we use a stripped-down version of our main optimization step. We first explain the main step of the algorithm, before detailing the two initialization stages.

### 4.1 Parameter Estimation

The quantities described in Section 3 can be divided into those related to the input data and those related to the underlying physics. In terms of unknowns, we distinguish between the parameters of the four trajectory segments of the two objects, which determine position and velocities, and the unknowns of the collision event. We will use a superscript $*$ to refer to the four trajectories, i.e., $* \in \{\text{pre-a, pre-b, post-a, post-b}\}$ and $\diamond$ for objects, i.e., $\diamond \in \{a, b\}$.

For each trajectory, we have the unknowns of each parabola. They are $b_1^*$, $b_2^*$, $\beta_{y0}^*$, while $\mathbf{b}_3^\diamond$ exist once for each body. Also, we have the global angles $\beta_x$, $\beta_{y1}$, and unknowns for angular momentum exist for each trajectory segment, $\mathbf{k}^*$. Additionally, the unknowns for the collision are: orientation for both objects at collision time $\mathbf{q}^{c,\diamond}$, location $\mathbf{x}^c$, impulse $j\mathbf{n}$, and the mass ratio $m^{b,a}$, as well as the coefficient of restitution $c$ and the time of the collision $t^c$.

Our optimization uses a dense set of 2D object centroids $\{\mathbf{p}_i^{2d}\}_{i=1}^N$ with depth values $\{d_i\}_{i=1}^N$ at times $\{t_i^{\text{obs}}\}_{i=1}^N$, and a sparse set of orientations for both bodies $\{\mathbf{q}_j^{\text{obs}}\}_{j=1}^M$ at times $\{t_j^{\text{obs}}\}_{j=1}^M$ from the input video. Later, in Section 4.2, we describe how to extract this data from the video. We use between 200 and 350 video frames with position information, and two semi-automatically annotated orientations per parabola segment, thus eight orientations in total. Additionally, we require a bounding box estimate for each body: $\mathbf{s}^\diamond$. Finally, we have a set of additional equations from the conservation laws, namely Equations (10-14) to tie the observations together according to the physical model. They represent our physics-based regularization encoded as additional equations.

As the number of equations from input data and physics is significantly larger than the number of unknowns, we have an over-determined problem. As the equations are non-linear, we compute a solution using non-linear least-squares. We summarize the unknowns in the vector $\theta$ for this least-squares optimization, and reformulate the equations to take the form

$$\underset{\theta}{\operatorname{argmin}} \sum_{\tau \in \texttt{Terms}} w^\tau f^\tau(\theta) \,, \tag{15}$$

where $w^\tau$ is weighting the corresponding energy term $f^\tau$, and $\texttt{Terms} = \{\texttt{mom}, \texttt{imp}, \texttt{g}, \texttt{ke}, \texttt{CoR}, \texttt{pxy}, \texttt{pz}, \texttt{ori}\}$. We now describe these individual energy terms.

**Respecting physics.** We directly turn the physics constraints in Equations (10-14) into energy terms by defining the least squares energy to be their residual. We combine the equations for momentum conservation (10,11) into $f^\texttt{mom}$:

$$f^\texttt{mom} = \frac{1}{2}\|\mathbf{v}^{\text{pre,a}} + \mathbf{v}^{\text{pre,b}} m^{\text{b,a}} - \mathbf{v}^{\text{post,a}} - \mathbf{v}^{\text{post,b}} m^{\text{b,a}}\|^2 \tag{16}$$

$$+ \frac{1}{2}\|\mathbf{p}^\text{a} \times \mathbf{v}^{\text{pre,a}} + \mathbf{k}^{\text{pre,a}} + m^{\text{b,a}} \mathbf{p}^\text{b} \times \mathbf{v}^{\text{pre,b}} + \mathbf{k}^{\text{pre,b}} -$$

$$\mathbf{p}^\text{a} \times \mathbf{v}^{\text{post,a}} - \mathbf{k}^{\text{post,a}} - m^{\text{b,a}} \mathbf{p}^\text{b} \times \mathbf{v}^{\text{post,b}} - \mathbf{k}^{\text{post,b}}\|^2 \,,$$

and the impulse equations (12,13) into $f^\texttt{imp}$:

$$f^\texttt{imp} = \frac{1}{4}\Big(\|\mathbf{v}^{\text{pre,a}} + \frac{j\mathbf{n}}{1} - \mathbf{v}^{\text{post,a}}\|^2 + \|\mathbf{v}^{\text{pre,b}} - \frac{j\mathbf{n}}{m^{\text{b,a}}} - \mathbf{v}^{\text{post,b}}\|^2$$

$$+ \|\mathbf{k}^{\text{pre,a}} + ((\mathbf{x}^\text{c} - \mathbf{b}_3^a) \times j\mathbf{n}) - \mathbf{k}^{\text{post,a}}\|^2$$

$$+ \|\mathbf{k}^{\text{pre,b}} - ((\mathbf{x}^\text{c} - \mathbf{b}_3^b) \times j\mathbf{n}) - \mathbf{k}^{\text{post,b}}\|^2\Big)\,. \tag{17}$$

Both terms are evaluated at collision time $t^\text{c}$, omitted for brevity. As our inputs are typically hand-held videos, we do not fully constrain gravity to act along the y-axis, but allow the optimization to slightly deviate, if necessary. For this we define the energy term

$$f^\texttt{g} = \Big(g - \begin{pmatrix}0 & 1 & 0\end{pmatrix} \mathbf{R}_{0,\beta_x,\beta_{y1}} \begin{pmatrix}0 & g & 0\end{pmatrix}^T\Big)^2\,. \tag{18}$$

Note that $\beta_{y0}$ is zero here, as this value represents rotation around the axis of gravity.

We formulate the kinetic energy limiting term from Eq. (9) as:

$$f^\texttt{ke} = \begin{cases}(E_k^\text{post} - E_k^\text{pre})^2 & \text{for } E_k^\text{post} > E_k^\text{pre} \\ 0 & \text{otherwise,}\end{cases} \tag{19}$$

where pre- and post-collision kinetic energies are evaluated in terms of velocities at the time of collision ($\times \in \{\text{pre}, \text{post}\}$):

$$E_k^\times = \mathbf{v}^{\times,\text{a}}(t^\text{c})^2 + \boldsymbol{\omega}^{\times,\text{a}}(t^\text{c})^T \mathbf{I}^\text{a}(t^\text{c}) \, \boldsymbol{\omega}^{\times,\text{a}}(t^\text{c})$$
$$+ m^{\text{b,a}} \mathbf{v}^{\times,\text{b}}(t^\text{c})^2 + \boldsymbol{\omega}^{\times,\text{b}}(t^\text{c})^T \mathbf{I}^\text{b}(t^\text{c}) \, \boldsymbol{\omega}^{\times,\text{b}}(t^\text{c})\,. \tag{20}$$

Finally, to disallow energy creation, we directly constrain the $c$ estimate to physical range of zero to one. For this we include a term

$$f^\texttt{CoR} = \begin{cases}c^2 & \text{for } c < 0 \\ (c-1)^2 & \text{for } c > 1 \\ 0 & \text{otherwise},\end{cases} \tag{21}$$

with $c$ calculated using Equation (14).

**Aligning to data.** Next, we formulate the data terms. Based on Eq. (5) we add terms for position fidelity. The notation is slightly complicated by the fact that we do not directly compute the difference of two 3D positions, but instead separately consider screen space and depth, each of which we will detail below.

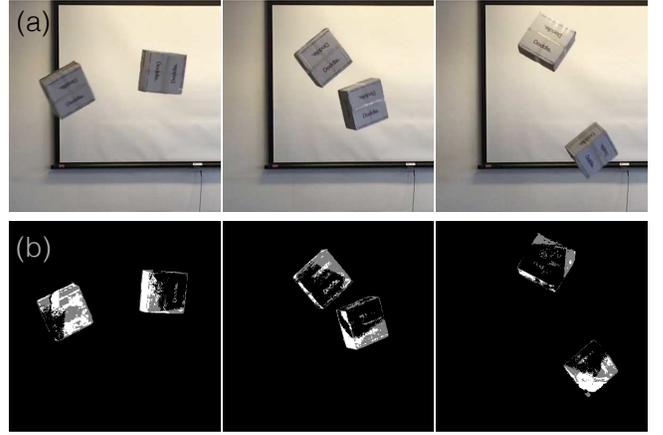

**Figure 4:** *In order to automatically retrieve measurements of the object centroids, we first post-process the input video (a) to subtract the background. This step yields approximate, and noisy image regions for the objects (b).*

For the screen space positions $\mathbf{p}_i^{2\text{d},*}$ (each at time $t_i^\text{obs}$) that were extracted from the video, we add an energy term penalizing screen-space deviations of the corresponding parabola with

$$\mathbf{p}_i^*(t_i^\text{obs}) = \begin{pmatrix} b_1^*(t_i^\text{obs} - t^\text{c}) \\ -\frac{g}{2}(t_i^\text{obs} - t^\text{c})^2 + b_2^*(t_i^\text{obs} - t^\text{c}) \\ 0 \end{pmatrix} + \mathbf{b}_3^\diamond$$

$$f^\texttt{pxy} = \frac{1}{N}\sum_{i=1}^N \|\mathbf{P}\,\mathbf{p}_i^*(t_i^\text{obs}) - \mathbf{p}_i^{2\text{d},*}\|^2 \,, \tag{22}$$

where $\mathbf{P}$ is the camera projection matrix, $N = |\{\mathbf{p}_i^{2\text{d},*}\}|$ and $*$ is again one of {pre-a, pre-b, post-a, post-b} depending on the input $i$, and $\diamond \in \{\text{a, b}\}$. Note that $\mathbf{b}_3^{\text{pre},\diamond} \equiv \mathbf{b}_3^{\text{post},\diamond}$ is shared among pre- and post-collision parabolas of the same object.

For the depth, we compute the energy terms with

$$f^\texttt{pz} = \frac{1}{N}\sum_{i=1}^N \|\begin{pmatrix}0 & 0 & 1\end{pmatrix}\mathbf{p}_i^*(t_i^\text{obs}) - d_i\|_\epsilon \,, \tag{23}$$

where $\|\cdot\|_\epsilon$ denotes a robust *Huber* [1964] loss function to allow for well-supported changes in depth. Specifically,

$$\|x\|_\epsilon = \begin{cases} x^2/(2\epsilon) & \text{if } x^2 \leq \epsilon \\ |x| - \epsilon/2 & \text{otherwise.}\end{cases} \tag{24}$$

The energy terms for the orientations are straight-forward to add based on Eq. (8):

$$f_j^\texttt{ori} = \frac{1}{M}\sum_{j=1}^M \|\mathbf{q}^*(t_j^\text{obs}) - \mathbf{q}_j^{\text{obs},*}\|^2 \,, \tag{25}$$

where $M = |\{\mathbf{q}_j^{\text{obs},*}\}|$, and $\mathbf{q}^*(t_j^\text{obs})$ is expressed as a series of explicit integration steps starting from $\mathbf{q}^{\text{c},\diamond}, \diamond \in \{\text{a, b}\}$. We use the analytic equations for cuboids to calculate the moment of inertia for our bodies. We pre-align their principal moments with the coordinate axes using SVD on the scanned models during a pre-computation step. We then scale it with the masses of the objects, *i.e.,* 1.0 and $m^{b,a}$. For a specific instance in time, $\mathbf{I}_0$ is transformed with the orientation at this time according to Equation (4).

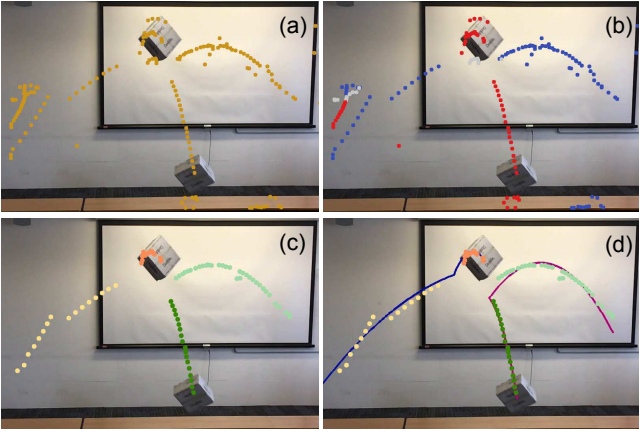

**Figure 5:** *The steps of our position initialization: (a) object region centroids; (b) initial labeling: pre-collision in blue, post-collision in red, and discarded points in gray; (c) re-assignment, and (d) final parabolas: object A in blue, object B in purple.*

**Collision point.** As our optimization has to navigate a highly non-linear energy landscape, a good initialization is crucial (see Sections 4.2 and 4.3). Additionally, we found that our optimization profits from constraining the collision point at first. We thus run our full optimization twice: first (Alg. 1, line 30) with a collision point $\mathbf{x}^c = (\mathbf{b}_3^a + \mathbf{b}_3^b)/2$, *i.e.,* constrained to lie at the midpoint of the collision positions of both objects. Removing these degrees of freedom at first increases robustness, especially for cases with lower quality data.

For the second run, we release the collision point to compute the final parameters of the collision (Alg. 1, line 31). While mass ratio, *c, etc.,* have been computed during the solve, we evaluate the final positions and orientations from our continuous, analytical parametrization.

In the following, we explain how to obtain measurements of screen space object centroids and 3D object orientations from the video.

### 4.2 Initializing Positions

We simultaneously estimate screen space object centroid measurements, assign them to trajectory segments, and initialize the collision in 3D. We assume that both objects are roughly equidistant to the camera at all times and that their 3D bounding box estimates from the initial scans are available. Essentially, we use the known gravitational acceleration and video frame rate to position the collision in 3D.

**Outline.** First, we calculate image region candidates for the reprojection of the objects in each frame. Then, we perform a RANSAC labeling involving several, modified parabola fits to compute temporally consistent assignments of the image region centroids to the four trajectory segments (see line 10 in Alg. 1) and to identify false-positive image regions. Finally, we iterate this process in a loop to determine the approximate time of collision.

**Image region calculation.** We estimate candidate center of mass positions in image space. We use a standard background subtraction pipeline consisting of Gaussian mixture modeling [Zivkovic 2004] and connected contour detection [Teh and Chin 1989]. Figure 4

**Algorithm 1:** SMASH

```
1  InitializePositions ()                    /* Sec. 4.2 */
2  InitializeOrientations ()                 /* Sec. 4.3 */
3  ReconstructCollision ()                   /* Sec. 4.1 */
4  Function InitializePositions              /* Sec. 4.2 */
5  Input  : video, s◇, P, f_fps, g
6  Assumption: β*_{y0} = 0, b^a_{3,z} ≡ b^b_{3,z}
7  t̃^c ← N/2
8  do
9  │  t^c_init ← t̃^c
10 │  for e=1 to 25 do                       /* RANSAC loop */
11 │  │  {p^{2d,*}_k}^{K=0.3N}_{k=1} ← AssignRnd(Sample({p^{2d}_i},0.3), t^c_init)
12 │  │  θ ← Solve({pxy, siz, g}, {}, {p^{2d,*}_k}^K_{k=1})
13 │  │  {p^{2d,*}_i}^N_{i=1} ← AssignGreedy({p^{2d}_i}, θ)
14 │  │  θ ← Solve({pxy, siz, g}, {}, {p^{2d,*}_i}^N_{i=1})
15 │  │  θ̂ ← Compare(θ̂, Score(θ, {p^{2d,*}_i}^N_{i=1}))
16 while t̃^c ≤ ⌊t^c_init⌋ or t̃^c ≥ ⌈t^c_init⌉
17 Output: {p^{2d,*}_i}^N_{i=1}, {d^*_i}^N_{i=1}, {t^{obs}_i}^N_{i=1}, t^c, b̃^*_1, b̃^*_2, b̃^◇_3, β̃^*_{y0}, β̃_x, β̃_{y1}
18 Function InitializeOrientations           /* Sec. 4.3 */
19 Input  : video, {p^{2d,*}_j}^M_{j=1}, {d^*_j}^M_{j=1}, {t^{obs}_j}^M_{j=1}, scan◇
20 Output: {q^{obs,*}_j}^M_{j=1} ← OpenDR(Input)
21 Function ReconstructCollision             /* Sec. 4.1 */
22 Input  : {p^{2d,*}_i}, {z^{2d,*}_i}, {q^{obs,*}_j}, {t^{obs}_i}, s◇, P, f_fps, g, t^c,
           b̃^*_1, b̃^*_2, b̃^◇_3, β̃^*_{y0}, β̃_x, β̃_{y1}
23 for ◇ ∈ {a, b} do
24 │  {j^{pre,◇}, j^{post,◇}} ← j | t^{obs,◇}_j closest before/after t^c
25 │  q^{c,◇} ← Slerp(q^{obs,◇}_{j^{pre,◇}}, q^{obs,◇}_{j^{post,◇}}, t^{obs,◇}_{j^{pre,◇}}, t^c)
26 │  for j^* ∈ {j^{pre,◇}, j^{post,◇}} do
27 │  │  (· ω̃^*)^T ← (2(q^{obs,*}_{j^*} − q^{c,◇})/(t^{obs}_{j^*} − t^c)) ⊗ (q^{c,◇})^{−1}
28 │  │  k^* ← R_{q^{c,◇}} I_0 R^{-1}_{q^{c,◇}} ω̃^*
29 x^c_init ← (b̃^a_3 + b̃^b_3)/2
30 θ ← Solve({mom, imp, g, ke, CoR, pxy, pz, ori}, {x^c})
31 θ ← Solve({mom, imp, g, ke, CoR, pxy, pz, ori}, {})
32 Output: b^*_1, b^*_2, b^◇_3, β^*_{y0}, β_x, β_{y1}, q^{c,◇}, k^*
33 Function Solve (Terms, FixedVars, Data = All)
34 return θ = solve Eq. (15) | Terms, FixedVars, Data
```

shows the output of the background subtraction step for three frames of a collision.

**RANSAC labeling.** We choose the middle of the video sequence as initial guess for the collision time $t^c$, more specifically, the midpoint in time between the two consecutive images. We then assign a randomly chosen centroid position from each video frame before and after $t^c$ to the pre- and post-collision parabolas respectively (line 11 in Alg. 1). We omit $f_{fps}/10$ frames (24 at 240fps) around collision time, as we found these to be particularly unreliable due to the merged silhouettes.

We perform a modified non-linear least squares fit to calculate the parabola parameters from Section 3.3 using the assigned screen space centroids, only considering the 2D position fidelity term (Equation (22)) and a new size penalty term:

$$f^{\text{siz}} = \frac{1}{N} \sum_{i=1}^{N} \left( \| \mathbf{P}(\mathbf{p}^*(t_i) + \frac{\mathbf{d}}{2}) - \mathbf{P}(\mathbf{p}^*(t_i) - \frac{\mathbf{d}}{2}) \| - s_i^{2d} \right)^2, \quad (26)$$

where $\mathbf{p}^*(t)$ is the position on a parabola at time $t$, $\mathbf{d}$ is the 3D

diagonal of the scanned object, $s^{2d}$ denotes the size of a circle fit to the image region corresponding to the selected centroid for this frame. This term pushes a parabola position towards a depth value so that the projected object diagonals match the image region sizes.

For this optimization, we share the z-coordinate of $\mathbf{b}_3$ between the two objects to ensure the objects meet in space. We also fix the parabola rotations around y to 0 ($\beta_{y0}^* = 0$) to stabilize the non-linear solve by constraining the space of solutions to be parallel to the camera plane. The subsequent main optimization step will refine these estimates.

The parabola fit allows us to greedily assign centroids to trajectory segments or label them as outliers (line 13 in Alg. 1).

To evaluate the quality of the obtained fit, we compute a score based on the residual of the parabola fit optimization, in conjunction with a SIFT feature distance $f^{\text{Im}}$ to measure the similarity of the selected image regions. For the former, we use the screen-space, area-weighted centroid-to-parabola distances $d_{\text{para}}$ from Eq. (22), multiplied with $f^{\text{inlier}}$, the ratio of centroids correctly explained (with an inlier distance equal to the median size of the identified regions). The image content score is calculated with a SIFT descriptor distance of the object regions between consecutive frames. The final score is weighted sum of these terms: $d_{\text{para}} f^{\text{inlier}} + w^{\text{Im}} f^{\text{Im}}$, see line 15 in Alg. 1. The fit with the lowest score is used to store the final set of screen-space centroids $\{\mathbf{p}_i^{2d,*}\}_{i=1}^N$, which are passed on to the main optimization stage. In addition to the 2D positions, we store the depth $\{d_i^*\}_{i=1}^N$ along each of the view rays for our depth terms (Eq. (23)), and their assignments to the trajectory segments. Figure 5 visually illustrates the steps of a single RANSAC iteration, typically converging in about 25 iterations.

**Collision time refinement.** Note that the optimization above computes an improved value for $t^c$ based on the intersection of the parabolas in Eq. (22). Initially, $t^c$ is set to the midpoint of a pair of video frames. If $t^c$ shifts before the first image of the pair, or to a point in time after the second one, we restart the calculation after selecting the previous, or next pair of frames, respectively. Their midpoint is the new initial guess for $t^c$, and we re-run the outlier detection, as described above. As this optimization runs very fast, we found that a sequential search for the correct collision time along the video works very well. Alternatively, a bisection search could be easily implemented.

### 4.3 Initializing Orientations

The previous initialization step yields a first set of 3D positions in time. We can render the scanned objects over the video on these trajectories to initialize the orientations, assuming that 3D scans of the filmed objects are available for this step. Using a differential renderer [Loper and Black 2014] and the object scans, we perform a first alignment of each object starting from 512 viewpoints uniformly sampled on a sphere. However, due to illumination variations we found the video information to be typically too ambiguous to automatically retrieve the correct orientations for all cases. Hence, we let a user check, and if necessary correct these alignments for two points in time for each parabola segment. This step takes less than a minute per collision sequence. An example of the generated candidate orientations can be seen in Figure 6.

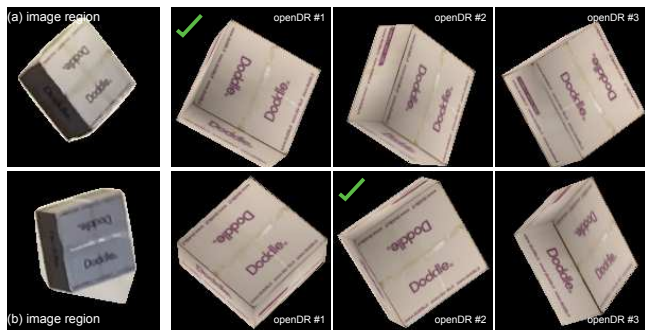

**Figure 6:** *Two examples of the image content (left) and the resulting candidate orientations (right). Green check marks highlight the selected orientations.*

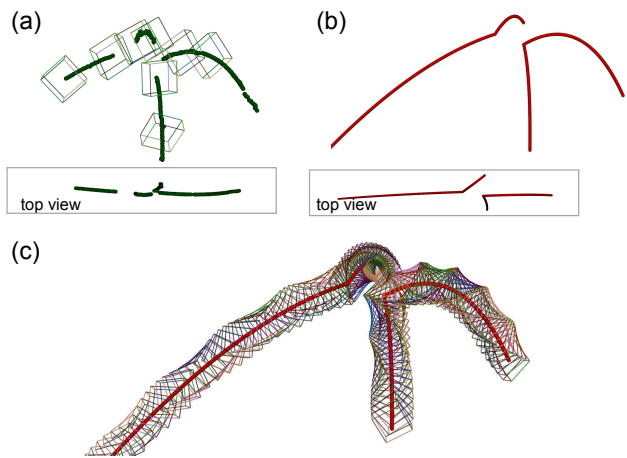

**Figure 7:** *a) the result of our initialization: labeled centroids and approximate depth values (green dots) together with the initialized orientations (green cuboids); b) 3D parabolas resulting from our full optimization (red curves), and c) motion over time (colored cuboids).*

The labelled screen-space centroid positions $\{\mathbf{p}_i^{2d,*}\}_{i=1}^N$ with depths $\{d_i^*\}_{i=1}^N$, and the $M = 8$ input orientations $\{\mathbf{q}_j^{\text{obs},*}\}_{j=1}^M$ are the input data for our full parameter estimation step (Section 4.1).

### 4.4 Implementation

We use CERES [Agarwal et al. 2016] for solving our non-linear least squares formulation, with sparse-Cholesky as linear solver type, and energy derivatives from automatic differentiation. Figure 7 shows an example result of the initialization, and the objects' recovered 3D positions and orientations.

Due to the non-linearity of the angular motion in Eq. (8) we typically use four integration steps per frame. Note that the sequence of integration steps is taken into account during the automatic differentiation performed in the solver. While higher-order integrators or even closed-form Poinsot solutions would be available to express a rigid body orientation over time, we found that explicit integration yields stable solutions in practice, and leave alternatives for future work. We use $f_{\text{fps}}$ to convert time $t$ to be measured in frames.

In our formulation above, we took care to minimize the number of free variables and energy terms. It turned out to be crucial for a good solution to use a formulation with shared variables, instead of enforcing equalities with highly weighted penalty terms. For example, instead of adding a term to minimize $\mathbf{b}_3^{\text{pre},a} - \mathbf{b}_3^{\text{post},a}$, we formulate the parabolas with a shared variable, which inherently enforces the equality of the offsets. Likewise, equivalent angles and orientations are implemented with shared variables instead of additional penalty terms. See Table 2 for runtimes. We have made code and data available online[1].

| Scene (#frames) | BgFg | IniPos | IniOri | Solve 1 | Solve 2 |
|---|---|---|---|---|---|
| box-1 (217) | 5.68s | 2m 6s | 18m 46s | 0.55s | 0.08s |
| box-2 (191) | 4.51s | 1m 51s | 17m 47s | 0.23s | 0.03s |
| box-3 (236) | 8.12s | 2m 0s | 19m 2s | 0.36s | 0.05s |
| helmet-duck (316) | 8.63s | 1m 40s | 24m 41s | 0.98s | 0.12s |
| duck-eleph. (204) | 5.53s | 1m 57s | 25m 55s | 3.08s | 0.18s |
| duck-rugby (316) | 11.86s | 1m 43s | 25m 11s | 0.65s | 0.36s |

**Table 2:** *Timings on a system with Intel i7-4700MQ and Nvidia Geforce GTX 770M. The second and third columns describe Section 4.2, the fourth column Section 4.3. The rightmost two columns reference lines 30 and 31 of Algorithm 1.*

## 5 Evaluation

We tested the robustness of our method and quantitatively evaluated the accuracy of the estimates. First, we report performance on synthetic data to test the system under different perturbations (Section 5.1) and then evaluate the system on a range of real-world sequences of increasing complexity (Section 5.2).

**Accuracy measures.** We evaluated the results as: (i) in a synthetic setup, we compare $c$ and $m^{b,a}$ estimates against ground truth values used to generate the test sequences; (ii) in real-world setups, we either compare with estimates obtained via other means (*e.g.,* using potential energy), or by evaluating consistency of the estimates across multiple collision sequences involving same object pairs. We use $\tilde{x}$ to denote our estimated value for quantity $x$, thus an output of the optimization. *E.g.,* $\tilde{c}$ for our estimate of a coefficient of restitution $c$. Where applicable, ground truth values *measured* by other means will be denoted with an $m$ subscript, *e.g.,* $c_m$.

**Weights.** In *all* the following tests we used a fixed set of weights: $w^{\text{mom}} = w^{\text{imp}} = 10$, $w^{\text{g}} = w^{\text{ke}} = w^{\text{CoR}} = w^{\text{ori}} = w^{\text{pxy}} = 1$, $w^{\text{pz}} = 0.1$, $\epsilon = 1$, and for the initialization phase $w^{\text{siz}} = 0.1$, $w^{\text{Im}} = 1/500$.

### 5.1 Validation on Synthetic Data

We tested our system using synthetic data obtained from forward simulations using the BULLET physics engine. We created a scene with two colliding cuboids under gravity and rendered the simulation as input video for our algorithm.

**(i) Robustness to initialization errors.** We separately tested the effect of errors in 2D centroid and orientation estimation in the initialization stage. We rendered the synthetic sequence at 60fps, initialized position for each time frame (Section 4.2), and added increasing amount of uniform noise to the retrieved 2D object centroids before fitting the parabolas. The algorithm provided robust estimates for both $c$ and $m^{b,a}$ up to perturbation margins of 30% (Figure 8(a)) with the noise margin being compared to the average screen-space size of the larger object. We ran 63 experiments with different noise margins ($0-30\%$) and found that our parabola fitting to be robust even under moderate errors on centroid estimates. For perturbations involving orientation annotations (Section 4.3), we added noise to each coordinate, and re-normalized the resultant quaternions. We observed robustness in both $\tilde{c}$ and $\tilde{m}^{b,a}$ up to perturbation of around $40°$ around actual orientation estimates (Figure 8(b)).

[1] http://geometry.cs.ucl.ac.uk/projects/2016/smash/

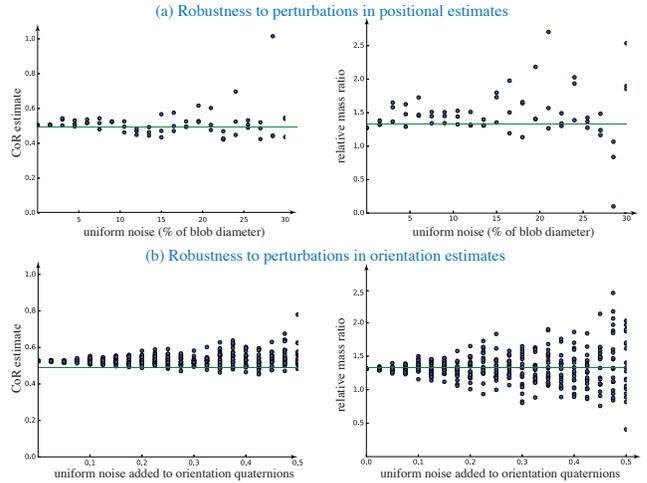

**Figure 8:** *Synthetic evaluation of robustness with increasing noise in the position and orientation estimates. Green lines denote ground truth, $c = 0.49$ and $m^{b,a} = 1.33$.*

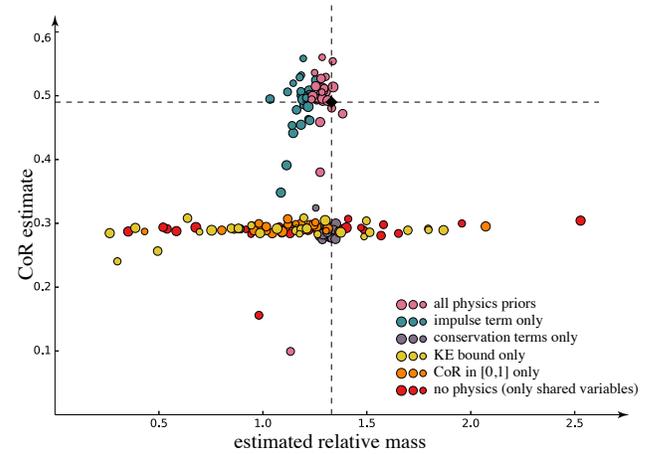

**Figure 9:** *Evaluation of the role of different terms. Impulse-based physics coupling is clearly the essential ingredient, while the other terms act as secondary regularizer. Larger circles show experiments with larger noise (positions and orientations). The input was a noise-free run of our automatic parabola initialization and synthetic orientations at least five timesteps away from collision time. Dashed lines denote ground truth, $c = 0.49$ and $m^{b,a} = 1.33$.*

**(ii) Effect of physics prior.** Next, in an important test, we assessed the contribution of the different physics-based priors in regularizing the optimization. We experimented by activating a selection of the optimization terms and evaluating estimation accuracy (see Figure 9). The impulse term has the single most important effect: while just based on the input video, the estimates for $c$ and $m^{b,a}$ deviate substantially, the impulse-based physics formulation provides valuable regularization resulting in reliable estimates even in the noisy setting. The secondary constraints in the form of bounds on $c$ and kinetic energy further improve the estimation accuracy.

**(iii) Effect of annotation frequency.** We measured the effect of fewer number of orientation annotations across time on estimated $m^{b,a}$ and $c$. Figure 11 shows the estimates for increasing annotation intervals. Note that even for the extreme case with an interval of around 20, *i.e.,* with only 4 annotations for a 90 frame sequence our solver computes reliable estimates. In all the other experiments, we used 2 orientation annotations for each of the 4 parabolas for a pair of colliding objects.

**(iv) Effect of friction.** Recall that in our formulation, we ignored the effect of friction. Hence, we evaluated the effect of ignoring friction on our estimates. In the experiments, we changed the (static) friction coefficient and coefficient of restitution of the simulator and re-ran the simulation. We then estimated $c$ and $m^{b,a}$ using our algorithm on the simulation output. As shown in Figure 10, we observed that while the relative mass estimates continued to be good, the coefficient of restitution was underestimated. This is explained by our model ignoring energy loss due to friction, especially when the friction coefficient is (synthetically) increased.

## 5.2 Testing on Real Sequences

We captured the real-world sequences with $1280 \times 720$ video with 120fps or 240fps using a regular iPhone6 with known camera intrinsics. The colliding objects were scanned in static positions using Kinect Fusion.

**Potential energy comparison.** One way to measure the coefficient of restitution is based on the ratio of energy loss during a collision. However, this requires a very controlled setting, where we can measure the change of energy using only the objects' potential energy, *i.e.,* when they have zero kinetic energy. We use different balls dropped from a controlled height $h$, measuring the apex of their trajectories after they bounce off the floor. When no energy is transferred into an angular motion, we can compute $c = \sqrt{h_{\text{after}}/h_{\text{before}}}$. We experimented using four different balls (ping-pong, tennis, football, and rugby) and computed the resulting $c$ values. We also used our algorithm to independently estimate $c$. As this setup involves one static object (the floor), we remove all unknowns of the second object from our solve. We took 3 sets of measurements for each object. Our estimates $\tilde{c}$ and the values measured with potential energy loss $c_m$ are listed in the table in Figure 12. Note that for the examples without any spin, the algorithm ran in an automatic mode as we do not require any orientation annotation.

Our method continues to consistently estimate $c$ even when some energy is stored/converted to kinetic energy throughout the recording, *i.e.,* the ball is spinning, or it starts off spinning (Figure 13). The simple estimation with potential energy is not valid for this case, while our formulation is still able to compute an estimate very close to the non-spinning case. Orientation annotations were provided for the scenes *rug-1* and *rug-2*. Please refer to the supplementary video for comparison.

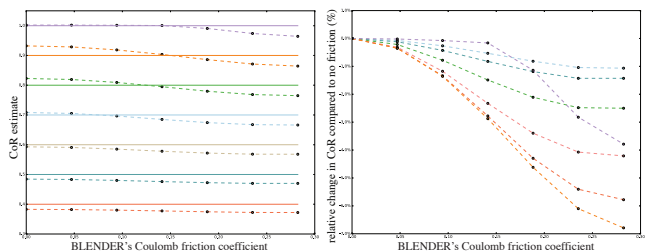

**Figure 10:** *Evaluation of the role of friction. We changed the friction coefficient and coefficient of restitution in the input simulation, and ran our algorithm on the simulator's output log (no noise). The c value estimates decrease a bit, which is expected, since some energy is lost to friction. Solid lines on left denote ground truth.*

## 5.3 Real-world Cases

**Collisions between box-like object pairs.** We next focus on collisions between pairs of freely moving objects. In contrast to the potential energy cases, there is no way to measure a ground truth value $c_m$ and we only present the value calculated with our method. We can however measure ground truth for the objects' mass ratio.

| scene | $\tilde{c}$ | $\tilde{m}^{b,a}$ | $m_m^{b,a}$ |
|---|---|---|---|
| box-1 | 0.28 | 0.91 | 1.0 |
| box-2 | 0.39 | 1.21 | 1.0 |
| box-3 | 0.20 | 1.0 | 1.0 |

**Table 3:** *Results of different collisions of cardboard boxes.*

Table 3 lists our estimates across three captured videos of the boxes colliding in different situations. Note that in all these examples, the video sequences around (estimated) collision times are ignored by our method as those frames did not provide any reliable position/orientation estimates due to overlapping objects. While the mass ratios are stable, and especially close to the ground truth for the last run, the $\tilde{c}$ vary more strongly. We attribute this partially to the fact that the cardboard boxes exhibit different stiffnesses depending on where they collide (based on the internal structure of the glued paper), in addition to the noise of the centroids estimated from the video. Figure 14(a)-(c) shows some frames from the box sequences.

**Effect of direct keyframe-interpolation.** We evaluate the effect of position and orientation interpolation while ignoring the underlying physics of collision. For this test, we used the initial position and orientation estimates obtained in our initialization stage as keyframes. As shown in Figure 15-middle, such a direct interpolation without access to physics-based constraints resulted in 'missing' the actual collision, *i.e.,* the duck and elephant 'separated' without actually colliding. In comparison, Figure 15-bottom shows that our method accurately reconstructs the collision.

**Collisions between non-box objects.** We next evaluate how our algorithm performs on collision videos involving non-box objects. We present three sequences of varying complexity: helmet-duck, duck-elephant, and duck-rugby. Table 4 lists estimation results.

Figure 14(d)-(e) show the first two results. As the frame-level comparison shows the reconstructed motion, both in terms of position and orientation, is visually very accurate. Although we validated the estimated mass ratios, we could not validate the accuracy of the estimated $\tilde{c}$ values. Qualitatively the $\tilde{c}$ estimates seem plausible: for example, the bike helmet gives a low estimate as it acts as a shock absorber; in contrast, the elephant and the rugby are bouncy and give high $\tilde{c}$ estimates, with the rugby giving the highest value.

| scene | $\tilde{c}$ | $\tilde{m}^{b,a}$ | $m_m^{b,a}$ |
|---|---|---|---|
| helmet-duck | 0.17 | 0.81 | 1.0 |
| duck-elephant | 0.46 | 2.33 | 2.0 |
| duck-rugby (Fig. 1) | 0.72 | 1.43 | 1.65 |

**Table 4:** *Estimates of internal physics properties ($\tilde{c}$ and $\tilde{m}^{b,a}$) of complex objects. The helmet clearly dissipates the most energy, whilst a sports ball is the most elastic by design.*

The duck-rugby sequence as shown in Figure 1 is the most challenging example as the objects remain hidden behind the curtain in the video for a large time interval. Still the reconstruction results closely follow the recorded video. Note that this example is particularly challenging since the duck is comparatively small and featureless, making the initial orientation estimates noisy.

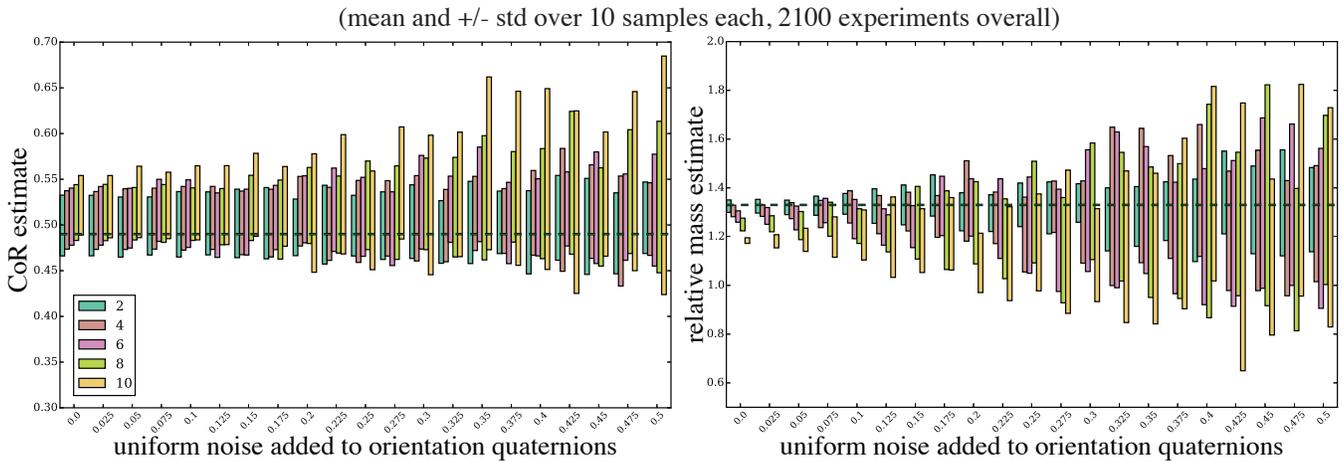

**Figure 11:** *Synthetic evaluation of robustness towards noise in the orientations. Gap (2, 4, 6, 8, 10) denotes the distance in time to the closest orientation around collision time. Dashed lines denote ground truth, $c = 0.49$ and $m^{b,a} = 1.33$.*

### 5.4 Authoring

We developed a BLENDER plugin (see Figure 16) to help author complex novel collision sequences using our parameter estimates from real collisions. The plugin allows the user to combine several observed collisions, time them to synchronize, or even stagger the interactions (*i.e.*, collisions). We used BULLET to obtain the object trajectories as accurately as possible using our estimated collision parameters. Note that the user does not have to guess physical parameters for objects (*i.e.*, $c$ and $m^{b,a}$ values). In Figure 16 we show the construction of a complex case of three pairs of rigid bodies colliding in mid air and two of them colliding again afterwards on their downward trajectories. Based on our reconstructed solutions, the new simulations behave naturally, and the objects collide as we would expect them to do from their real-world counterparts.

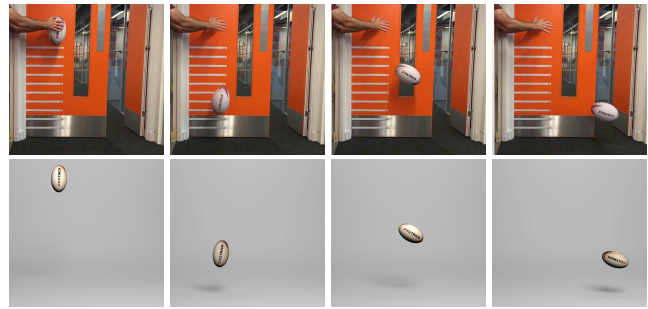

**Figure 13:** *Rotating rugby ball scene. Estimated $c$ using our method is $\tilde{c} = 0.78$, matching no-spin measurement $0.8$ as detailed in Fig. 12.*

### 5.5 Limitations and Discussion

A key limitation of our method is that it only applies to rigid body collisions. In case of deformable bodies, the physical models we employ can be unsuitable. Empirically, however, we observed that our estimates are still reasonable for near rigid collisions (see Figure 17). Our algorithm in its current form can reconstruct real-world effects under the assumption of zero friction, but including a friction model in our optimization could potentially increase the accuracy of our reconstructions.

While our current pipeline still requires user input (2 orientations per parabola segment), it is worth pointing out that the user-annotated orientations represent only a small part of the overall input data. All the 3D positions of the objects are reconstructed fully automatically. The user input is typically only required to refine the orientations, which are important for angular motions, and thus help to retrieve the physical parameters.

Also, as we only rely on a single input video, we need enough visual information about the collision event to be visible. For example, if a collision is happening purely along a view direction ray, our method will most likely underestimate the objects' velocities. Likewise, objects that move very quickly might not exhibit enough change of direction from gravity. In such a case, our method requires more accurate centroids to reliably recover a depth estimate from gravity.

Lastly, it is noticeable in some reconstructions (such as Figure 14a) that the objects intersect each other at collision time. This can happen, as our optimization does not include any explicit collision detection or resolution terms. While this might look like a limitation at first, we believe the accurate contacts of the other scenes actually point to a very high accuracy of the reconstructions, which was achieved without taking the exact shapes of the objects into account.

| scene | $\tilde{c}$ | $c_m$ |
|---|---|---|
| pong-0 | 0.55 | |
| pong-1 | 0.57 | 0.54 |
| pong-2 | 0.55 | |
| tennis-0 | 0.71 | |
| tennis-1 | 0.73 | 0.76 |
| tennis-2 | 0.71 | |
| football-0 | 0.66 | |
| football-1 | 0.67 | 0.77 |
| football-2 | 0.67 | |
| rug-0 | 0.80 | 0.82 |
| rug-1 (spin) | 0.80 | - |
| rug-2 (spin) | 0.78 | - |

**Figure 12:** *Various $c$ estimates obtained by our method as compared against $c_m$ obtained using alternate potential energy-based estimation, which is valid only when the object has no angular momentum. All the examples are instances of collision of a single object against an infinite mass object (i.e., carpeted ground).*

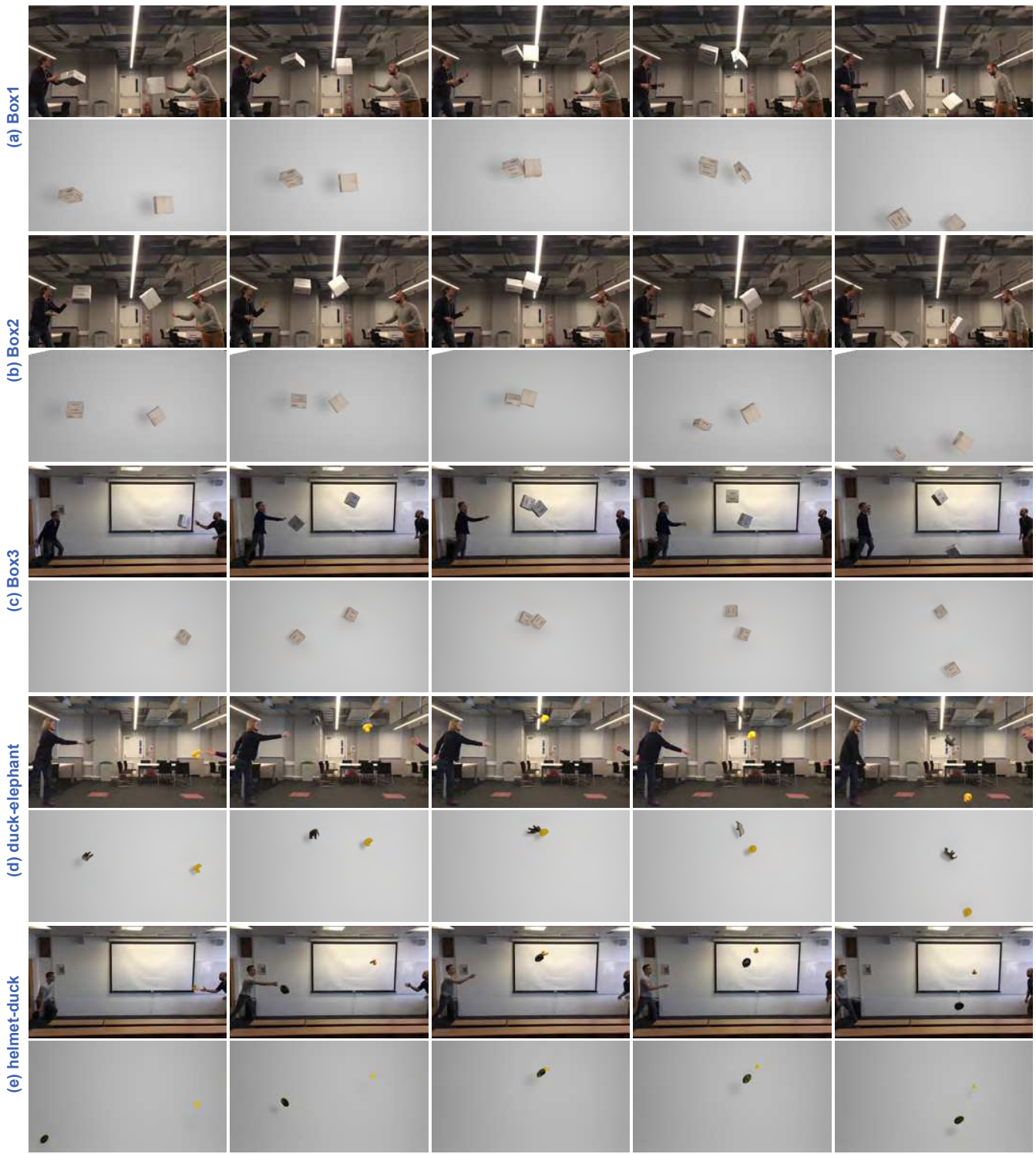

**Figure 14:** *We tested our algorithm on multiple collision sequences. Here we show a selection of images of input frames and corresponding reconstructed frames. Please refer to the supplementary video.*

## 6 Conclusion

We presented SMASH, a data-driven framework for capturing, reconstructing, and authoring collision sequences. The key idea is to regularize the problem of reconstructing collisions of object pairs from videos using laws of rigid body physics. We demonstrate how to reconstruct plausible collision sequences in 3D, by observing objects in motion away from the collision instant. The method outputs physical parameters that would otherwise be very difficult to acquire, *e.g.,* the coefficient of restitution and the mass ratio of the participating objects. The information can then be readily used to author new collision sequences.

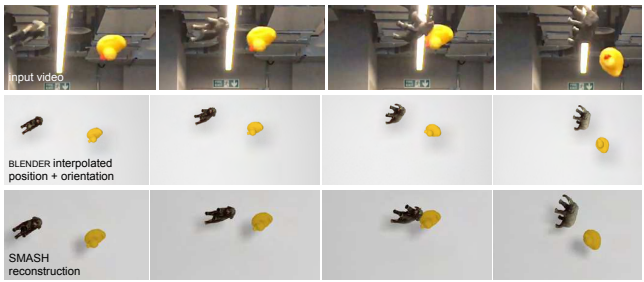

**Figure 15:** *Direct keyframe interpolation (middle) versus our reconstructed result (bottom). Note that the interpolated sequences completely miss the collision (middle row, third from left) as it remains oblivious to any physics considerations.*

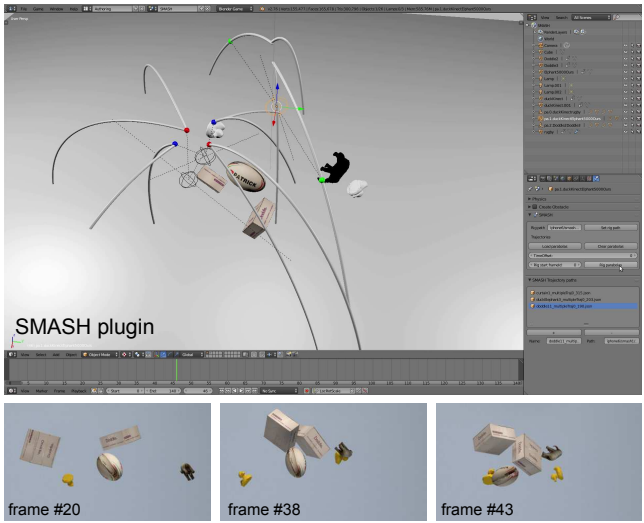

**Figure 16:** *Our SMASH plugin (top) can be used to author and synchronize collisions using the recovered parameters from our reconstruction. Trajectories, mass ratios and c values are read in from the optimization automatically to produce a realistic collision sequence (bottom). Please refer to the supplementary video.*

While we presented a first workflow for reconstructing collision sequences, many avenues for improvements remain. Our current implementation still expects user guidance to disambiguate the objects' orientations. It will be an interesting avenue for future work to iteratively apply our full pipeline to automate this step. Potentially, this could also be used to approximate the shape of the object, and thus make the initial scanning step unnecessary.

Finally, we believe our method only represents a small step towards a tighter integration of video analysis, dynamic geometry acquisition, and physics simulations. There are many other physical phenomena, such as deformable objects or fluids, that could be included in order to on one hand better understand what is happening in a recorded event, and on the other hand reconstruct the underlying physics for computer animation purposes.

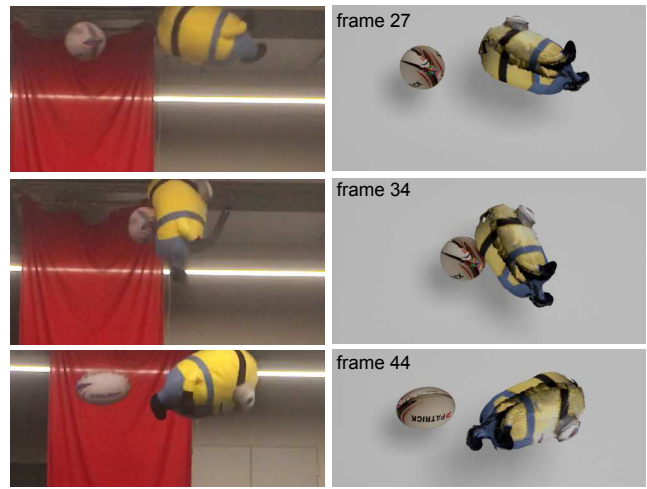

**Figure 17:** *Although designed for rigid collisions, we observed that our method still can produce reasonable estimates for near rigid collisions as shown by the minion-rugby sequence. However, the reconstruction is inaccurate around the collision time (frame #34).*

## Acknowledgements

We thank our reviewers for their invaluable comments. We also thank Gabriel Brostow, Paul Guerrero, Martin Kilian, Oisin Mac Aodha, Moos Hueting, James Hennessey, Stephan Garbin, Chris Russell, Clément Godard, Peter Hedman and Veronika Benis for their help, comments and ideas. This work was partially funded by the ERC Starting Grant SmartGeometry (StG-2013-335373), the ERC Starting Grant realFlow (StG-2015-637014), UCL Impact, a Google PhD Fellowship, and gifts by Adobe.